\documentstyle[11pt,aaspp4]{article}
\begin{document}
\title{HST/STIS UV Spectroscopy of Two Quiescent X-ray Novae:
\newline A0620-00 and Centaurus X-4}

\author{Jeffrey E. McClintock}
\affil{Harvard-Smithsonian Center for Astrophysics}
\affil{60 Garden St., Cambridge, MA 02138}
\affil{jem@cfa.harvard.edu}
\and
\author{Ronald A. Remillard}
\affil{Center for Space Research, MIT}
\affil{Cambridge, MA 02139}
\affil{rr@space.mit.edu}

\singlespace

\begin{abstract} In 1998 we made UV spectroscopic observations with
HST/STIS of A0620-00 and Cen~X-4, which are two X-ray novae (aka soft
X-ray transients). These binary systems are similar in all respects
except that the former contains a black hole and the latter contains a
neutron star.  A UV spectrum (1700-3100\AA) is presented for the
quiescent state of each system in the context of previously published
UV/optical and X-ray data.  The non-stellar, continuum spectrum of
black hole A0620-00 has a prominent UV/optical peak centered at
$\sim$~3500\AA.  In contrast the spectrum of neutron-star Cen X-4
lacks a peak and rises steadily with frequency over the entire
UV/optical band.  In the optical, the two systems are comparably
luminous.  However, black hole A0620-00 is $\sim 6$ times less
luminous at 1700\AA, and $\sim 40$ times less luminous in the X-ray
band. The broadband spectrum of A0620-00 is discussed in terms of the
advection-dominated accretion flow model. \end{abstract}

%\begin{slocitlum}{2}

\footnotesize
\noindent \footnoterule
\noindent
$^{1}$Based on observations with the NASA/ESA {\it Hubble Space Telescope}
obtained at the Space Telescope Science Institute, which is operated by
the Association of Universities for Research in Astronomy, Inc., under
NASA contract NAS5-26555.

\normalsize
\section{Introduction}

In the fall of 1975, the X-ray nova A0620-00 reached an intensity of
50 Crab to become the brightest (non-solar) celestial X-ray source
that has ever been observed.  After the system returned to quiescence
in the fall of 1976, it was found to be a 7.8 hr binary containing a
K5 dwarf secondary and a non-stellar continuum source, which was
attributed to an accretion disk (Oke 1977; McClintock et al. 1983).
The brightness of the quiescent optical counterpart (V = 18.3) and the
low mass of the secondary motivated a radial velocity study which led
to the discovery of the first black-hole primary in an X-ray nova
(McClintock \& Remillard 1986).

Already in 1981, the quiescent X-ray source was known to be
extraordinarily faint with L$_{x} < 10^{-6}$L$_{max}$ (Long et
al. 1981). McClintock (1986) noted that this faintnesss posed the
following puzzle: By close analogy with dwarf novae, A0620-00's
quiescent accretion disk (M$_{V} \sim 7$) implied a mass accretion
rate of \.{M} $\sim 10^{-11}$~M$_{\odot}$ yr$^{-1}$. Thus, if one
assumed the canonical efficiency of $\sim$10\% for converting
gravitational energy into radiant energy, then one would expect the
source to be $\sim$1000 brighter in X-rays than was observed.

A few explanations for the lack of X-rays were suggested (e.g. de Kool
1988; Huang and Wheeler 1989), but none was satisfactory. Moreover,
these suggestions were overtaken by the advection-dominated accretion
flow (ADAF) model (Narayan \& Yi 1994, 1995, and references therein).
 An early application of the ADAF model by Narayan, McClintock, and Yi
(1996) showed that it could account naturally for both the broadband
spectrum of A0620-00 and for its minuscule X-ray luminosity of
$\approx 6 \times 10^{30}$ ergs~s$^{-1}$ (McClintock, Horne \&
Remillard 1995; hereafter MHR95).

Similar fits were made successfully to the broadband spectra of other
quiescent X-ray novae (Narayan, Barret, \& McClintock 1997a; Hameury
et al. 1997). The results presented herein are discussed in terms of
these spectra and the ADAF model, which is based on the idea that at
low \.{M} the gravitational potential energy released in an accretion
flow may be stored as thermal energy rather than being radiated. An
ADAF flow is optically thin, nearly radial, and extraordinarily hot
near the compact object. Thus in the case of a black hole, nearly all
of the gravitational energy can be advected through the event horizon
and swallowed by the hole, thereby accounting for the tiny X-ray
luminosity of A0620-00 quoted above.  On the other hand, a neutron
star accreting via an ADAF is expected to be much more luminous for
the same \.{M} because the accreting gas will strike the surface and
heat the star (assuming it is not deflected by the propeller effect),
thereby converting gravitational energy into radiation with the
canonical efficiency of $\sim 10\%$ (Narayan, Garcia, \& McClintock
1997b; Menou et al. 1999).

We reported earlier on a UV spectrum of A0620-00 of modest quality
that was obtained with the FOS during HST Cycle 1 (MHR95).
Here we report on UV spectra of both A0620-00 and Cen~X-4 that are
of excellent quality and were obtained in 1998 with the STIS during
Cycle 7. The UV spectrum of Cen~X-4 in quiescence is the first such
spectrum of this object to be published.  

Cen~X-4 is a type I burst source and therefore contains a neutron
star primary, which is only about 15\% as massive as the black hole
primary in A0620-00 (Shahbaz, Naylor, \& Charles 1994).  The
quiescent X-ray luminosity of Cen~X-4, L$_{x} \sim 2.5 \times
10^{32}$ ergs~ s$^{-1}$, is $\sim 40$ times larger than the X-ray
luminosity of A0620-00 (Asai et al. 1998; MHR95). In most other
respects, however, Cen~X-4 and A0620-00 are very similar. For
example, the orbital period of Cen~X-4 is 15.1 hrs {\it vs.} 7.8 hrs
for A0620-00, and so their predicted mass transfer rates are quite
comparable (Menou et al. 1999). Their estimated distances are about
the same: 1.2 kpc (Chevalier et al. 1989; Barret, McClintock, \&
Grindlay 1996).  Both systems contain K-dwarf secondaries, and both
are known to have erupted twice: Cen~X-4 in 1969 and 1979, and
A0620-00 in 1917 and 1975.

\section{Observations and Analysis}

We report on eight UV spectra of A0620-00 (V616~Mon) and one UV
spectrum of Cen~X-4 (V822~Cen) that were obtained with the HST/STIS
spectrograph.  Some details of the observations are presented in
Table~1.  All of the observations employed the G230L grating, the
clear filter, the 52''x0.5'' aperture and the NUV-MAMA detector. The
spectral resolution is 3.2\AA\ (FWHM).  The data were recorded in the
``time-tagged" mode.

Our results presented herein are based on the calibrated ``pipeline''
spectral data provided to us by STScI.  We converted each of the
``1-D" table files into several 1-D IRAF spectra containing the flux,
the error in the flux, the background, etc.  The first UV spectrum
obtained for A0620-00 (entry~2 in Table~1) has a high background level
and we rejected it; no problems were found with the remaining
spectra. Thus, the single spectrum of A0620-00 presented herein is a
sum of seven individual spectra (entries~3-6 and 9-11 in Table~1) with
a total exposure time of 4.8 hours.  Only a single spectrum was
obtained of Cen~X-4 (Table~1, entry~8); the exposure time was 0.6
hours.

The fluxes were dereddened using the interstellar extinction law of
Cardelli, Clayton \& Mathis (1989).  For A0620-00 we adopted a
reddening of E(B-V) = 0.35 (Wu et al. 1983; also, see Oke \&
Greenstein 1977).  For Cen~X-4 we used E(B-V) = 0.10 (Blair et
al. 1984).  The uncertainty in the reddening of either source probably
does not exceed 0.05 mag (Wu et al. 1976, 1983; Blair et al. 1984). In
$\S3.4$ we illustrate the effects of this uncertainty on our UV
spectra.

To confirm that A0620-00 and Cen~X-4 were quiescent near the time of
the STIS observations, we obtained V-band CCD images of both fields
using the FLWO 1.2m telescope at Mt. Hopkins, AZ.  As Table~1 shows,
these optical observations occurred about 11 hours before the STIS
observation of Cen~X-4 and about 25 hours before the longer sequence
of observations of A0620-00.  The airmass and seeing were,
respectively, 1.3 and 2.5" for the observations of A0620-00 and 2.2
and 2" for CenX-4. The sky was clear during the observations of both
objects. All of the data were reduced in the standard way using bias
frames, dome flats and the NOAO IRAF reduction package.  The absolute
calibration for A0620-00 was derived from Landolt (1992) stars in four
fields, which were observed throughout the night.  The absolute
calibration for Cen~X-4 was obtained from the magnitudes of two nearby
comparison stars (Chevalier et al. 1989).

We also present new results on the V-band optical variability of
A0620-00 based on the following data collected by us: (1) One 20-min
CCD image was obtained on 1992 February 5.156 UT using the McGraw-Hill
1.3m telescope; (2) one 10-min image was obtained on 1992 April 7.006
UT using the CTIO 1.5m telescope; and (3) 23 images were obtained
during four nights--1995 December 21 and 1996 January 13-15--using the
FLWO 1.2m telescope. In all cases, the data reduction and analysis
were performed using IRAF.  The absolute calibration relies on the
magnitudes of three nearby field stars, which were derived from the
1998 March~3 observation of A0620-00 described above.

\section{Results}

\subsection{Spectra of A0620-00 and Cen~X-4}

The spectra of A0620-00 and Cen~X-4 are shown in Figure~1ab.  No
correction has been applied for interstellar reddening. The prominent
emission feature in both spectra is Mg~II
$\lambda\lambda2796,2803$. For A0620-00, the line has the same
equivalent width (EW) on 1998 March~4 and on May~5 (Table~1): $174 \pm
8$\AA.  Similarly, the line width is the same for both observations of
A0620-00: $23 \pm 2$\AA\ (FWHM), where a small correction has been
applied for the 3.2\AA\ spectral resolution of the STIS.  For Cen~X-4,
the EW of the Mg~II line is $63 \pm 5$\AA\ and the line width
corrected for instrumental resolution is $13.5 \pm 2.0$\AA\ (FWHM).
Another significant emission feature, which is also present in both
spectra, is Fe~II $\lambda\lambda2586-2631$. Both the Mg~II and Fe~II
emission lines appear with comparable EWs in the 1992 spectrum of
A0620-00 (MHR95).

The continuum for A0620-00 is faint and steadily decreases with
wavelength to f$_{\lambda} \lesssim 6 \times 10^{-18}$ ergs~s$^{-1}$
cm$^{-2}$ \AA$^{-1}$ for $\lambda \lesssim 2300$\AA.  The faint
continuum is due partly to the cutoff in the source spectrum at short
wavelengths (as illustrated below), and partly to extinction (E$_{B-V}
= 0.35$).  On the other hand, the observed continuum spectrum of
Cen~X-4 is relatively flat over the entire range and actually rises
somewhat with decreasing wavelength (Fig.~1b).

\subsection{Optical and X-ray Variability of Cen~X-4 and A0620-00}

The quiescent optical counterparts of both A0620-00 and Cen~X-4 are
modulated at their orbital periods with semiamplitudes of $\approx
0.1$ mag.  The mean brightness of Cen~X-4, however, varies by much more
than this. Chevalier et al. (1989) report on 570 V-band measurements
obtained over a 3-year period.  During most observing runs they found
V $\approx 18.6-18.2$. However, they frequently found that the
counterpart brightened further to V $\approx 18.1-17.7$, which they
call the ``active state.''  Occasionally, the variations occurred from
night to night, but usually they occurred on a longer time scale.
Similar variability in B and B-V was extensively observed by Cowley et
al. (1988). Furthermore, despite the paucity of X-ray measurements,
the quiescent X-ray flux of Cen~X-4 has been observed to vary by a
factor of 3 in less than four days (Campana et al. 1997).  Thus, both
the V-band and X-ray intensity of Cen~X-4 have been observed to vary
by a factor of $\sim 2-3$ on time scales of a day or longer. Our
V-band observation of Cen~X-4, which was performed shortly before our
STIS observation (Table~1, entries~7-8), gave V = $18.2 \pm 0.1$,
which is consistent with values reported for the quiescent state
(e.g. V = 18.6-18.2; Chevalier et al. 1989).

No comparable degree of variability has been reported for A0620-00 in
quiescence.  The earliest value for the quiescent V magnitude is given
by Oke (1977): V = 18.35 on 1976 November 15-17.  Here we report in
chronological order a number of new results (see $\S2$): (1) V = 18.20
$\pm 0.05$ on 1992 February 5.156 UT; (2) V = 18.20 $\pm 0.05$ on 1992
April 7.006 UT; (3) $\overline{V} = 18.25 \pm 0.08$ (n = 23) on 1995
December 21 and 1996 January 13-15, where the uncertainty is the
sample standard deviation, and the mean magnitude is consistent with
the mean magnitudes computed for each of the four individual nights;
and (4) V = 18.37 $\pm$ 0.05 on 1998 March~3 (see Table~1).  Thus,
apart from orbital modulation, the mean magnitude of A0620-00 appears
to be relatively stable: $\overline{V} \approx 18.3$.  There is no
useful data on possible X-ray variability.  Therefore, based on the
very limited data available for A0620-00 in quiescence, we tentatively
conclude that the mean magnitude of its optical counterpart is
significantly less variable ($\overline{\Delta V} \approx 0.1$ mag)
than the quiescent counterpart of Cen~X-4 ($\overline{\Delta V}
\approx$ 1 mag).

\subsection{Ultraviolet Variability and Reddening of A0620-00}

The STIS spectra of A0620-00 obtained two months apart provide strong
evidence for UV variability.  The March~4 and May~5 spectra (Table~1),
excluding the Mg~II line, are compared in Figure~2.  The STIS flux
value in a particular 100\AA\ band is the mean of three flux
measurements for the May~5 data and the mean of four measurements for
the March~4 data (see Table~1). The uncertainty in a given band is the
sample standard deviation of the (3 or 4) flux measurements in that
band.  Here and elsewhere we use the standard deviation because it
provides a realistic estimate of the random error which includes the
effects of source variability and counting statistics.

The uncertainties shown in Figure~2 (for 2400\AA\ $\leq \lambda \leq
3000$\AA) are typically 12\%, whereas the uncertainties due to
counting statistics are much smaller ($\sim$ 3\%). This implies that
A0620-00 varies by $\sim$ 12\% on a time scale of one 90-min HST
orbit. (For evidence of variability on a time scale of minutes, see
$\S2.2$ in MHR95.)

Comparing the nominal spectra plotted in Figure~2, we conclude that
the continuum flux ($2250 < \lambda < 3150$\AA) decreased by a factor
of $\approx 1.25$ in 1998 between March~4 and May~5. We have ignored
the 4\% uncertainty in the calibration and stability of the STIS with
its MAMA detector (STIS Instrument Handbook, Vers. 2.0).

Given the significant reddening of A0620-00, E(B-V)~$\approx~0.35$
mag, we examined the spectra for evidence of absorption near 2200\AA.
The March 4 spectrum shown in Figure 2 contains such a feature.  We
examined the significance of this broad feature by binning the data in
50\AA\ intervals and restricting our attention to the wavelength range
1900-2600\AA.  We corrected this spectrum for increasing amounts of
IS reddening corresponding to E(B-V) = 0.1, 0.2, 0.3, etc. (Cardelli
et al. 1989).  For each of these data sets, the uncertainties in the
fluxes were represented by the sample standard deviation, as in
Figure~2.  In the usual way, we fitted the corrected data sets to
power law spectra (F$_{\lambda}~\sim~\lambda^{n}$) and computed
$\chi^{2}$~{\it vs.}~E(B-V).  Acceptable fits (i.e. data sets with
$\chi^{2}~\sim1$ and a weak absorption feature) were obtained for
$0.3~\lesssim$~E(B-V)~$\lesssim 0.7$, with a best value near 0.5.
This value is consistent with the one determined by Wu et
al. (1976,1983) during the 1975 outburst of A0620-00 (see $\S2$);
however, our value is much less precise due to the variability and
faintness of the quiescent source.  For the fainter May 5 spectrum
(Fig. 2), no 2200\AA\ feature is evident; in fact, no significant flux
was detected for $\lambda \leq$ 2200\AA.

\subsection{A Comparison of the UV/Optical Spectra of A0620-00 and Cen~X-4}

The spectra of the two sources corrected for reddening are compared in
Figure~3ab. To facilitate our analysis of the STIS continuum spectra,
we have clipped out the Mg~II and Fe~II emission lines and averaged
the spectra in 100\AA\ intervals.  Reddening corrections have been
applied and the results are expressed in units of log($\nu$F$_{\nu}$)
{\it vs.} log($\nu$), which have been used extensively in modeling the
spectra of X-ray novae (e.g. Narayan et al. 1997a).  The STIS fluxes
and the V-band fluxes reported here are plotted as filled circles. The
effects on the STIS spectra of varying the reddening by $\pm~0.05$ mag
from the nominal values are indicated by the flanking histograms.  In
the spectrum of A0620-00 (Fig.~3a), FOS UV/optical fluxes obtained six
years earlier and some additional optical data for $\lambda \geq$
5000\AA\ are plotted as open squares (see Narayan et al. 1996, and
references therein).  All of the longer wavelength flux data ($\lambda
\gtrsim$ 3500\AA) shown here is non-stellar: i.e. the data have been
corrected by subtracting an approximate contribution due to the K-dwarf
secondary (see below).

For A0620-00 (Fig.~2a), a factor of 2.0 discrepancy in flux level is
evident in the 2400-3000\AA\ overlap range between the 1992 FOS
spectrum and the 1998 STIS spectrum.  As noted in $\S3.3$, the
absolute photometric accuracy of the STIS data are 4\%.  Thus the only
simple explanations for the offset between the spectra are (1) the FOS
fluxes are in error, (2) the source varied or (3) both 1 \& 2 are
true.  We cannot decide this question with the available information.
However, we note that the FOS sensitivity in Cycle 1 was handicapped
by the pre-COSTAR optics.  Furthermore, the source was faint and the
FOS provided no direct measure of the background (MHR95). There is one
quantitative point that narrows the FOS/STIS discrepancy: In 1992 we
observed A0620-00 with two FOS dispersers and found that the ``blue
prism'' fluxes (which we adopted in MHR95 and are shown plotted here
in Fig.~3a) exceeded the G160L grating fluxes by $\sim 35$\% in the
2200-2400\AA\ overlap range (MHR95). Had we adopted the grating fluxes
(now considered more reliable), the FOS/STIS discrepancy would be a
factor of 1.5 instead of 2.0 (as shown in Fig.~3a).  On the other
hand, source variability may be largely responsible for the
discrepancy.  Although very little V-band variability has been
observed over 12 years ($\S3.2$), the STIS results do provide strong
evidence for a $\sim 25\%$ variation in the UV flux in two months
($\S3.3$ and Fig.~2).

The non-stellar spectrum of A0620-00 at longer wavelengths ($\lambda
\gtrsim$ 3500\AA) is somewhat uncertain because one must subtract a
large stellar component due to the K-dwarf secondary.  Subtracting the
spectrum of a K5V field star, Oke (1977) first derived the non-stellar
spectrum of A0620-00 (3200-10000\AA).  Specifically, he found that the
non-stellar component contributed 43 $\pm$ 6\% of the total light in
the V-band.  We also used a K5 dwarf and a somewhat different
technique to obtain very similar results at six later epochs (MHR95).
On the other hand, using a K3/4 dwarf, Marsh, Robinson \& Wood (1994)
found a much smaller contribution for the non-stellar component: 17
$\pm$ 4\% at $\sim$4800\AA\ and 6 $\pm$ 3\% at $\sim$6300\AA. In
Figure~2a, the non-stellar optical spectrum based on Oke's
decomposition is plotted as open squares.  The dashed line indicates
approximately the effect on Oke's spectrum of applying the results of
Marsh et al. (1994), which call for a $\sim$3-fold decrease in the
non-stellar spectrum at 4800\AA\ and a $\sim$6-fold decrease at
6300\AA, as indicated by the pair of open triangles in Fig.~3a. The
difference between the two optical spectra is most likely due to the
choice of a proxy star to represent the secondary of A0620-00.  Marsh
et al. chose a hotter star with weaker lines (K3/4 {\it vs.}  K5),
which would be expected to give a smaller non-stellar contribution.
In any case, we tentatively conclude that the optical spectrum of the
non-thermal radiation is relatively uncertain, but it is very probably
bounded by the open squares and dashed line in Fig.~3a.

For A0620-00, the total V magnitude was the same for Oke in 1976 as it
was for us in 1998 (see $\S3.2$).  Thus in plotting this data point in
Figure~3a, we have assumed that the fraction of non-stellar light is
43\%, as derived by Oke (1977).  For Cen~X-4 at V (Fig.~3b), we have
assumed nominally that 30\% of the total light (corresponding to V =
18.2; see $\S3.2$) is non-stellar; the liberal error bar assumes that
the non-stellar fraction is within the range 0.25-0.40 (Chevalier et
al. 1989; McClintock \& Remillard 1990; Shahbaz, Naylor \& Charles
1993).
 
The key result of Figure~3ab comes from comparing the reliable STIS
spectra of A0620-00 and Cen~X-4 in the range 1800-3100\AA.  The
absolute photometric accuracy of these spectra is 4\% ($\S3.3$), and
they are not at all affected by ``light pollution'' from the cool
secondary stars.  {\it These spectra are strikingly different: The
intensity of A0620-00 decreases by a factor of $\sim 3$ in the
interval 1900-3000\AA\ ($15.0 <$ log($\nu$) $< 15.2$). In contrast,
the intensity of Cen~X-4 rises with frequency over the entire
observed UV band.}

\subsection{The Broadband Spectra of A0620-00 and Cen~X-4 Compared}

The spectra of both sources are shown in Figure~4ab extending from the
optical to the X-ray. The UV/optical spectrum is the same as shown in
Figure~3ab.  The upper limit near 50~eV (log$(\nu) \sim 16.6$) for
A0620-00 is deduced from an upper limit on HeII $\lambda4686$ emission
(Marsh et al. 1994; Narayan et al. 1996). For both sources the X-ray
data near 5 keV was obtained with ASCA (Asai et al. 1998).  The
detection plotted for A0620-00 at 1 keV was obtained with the ROSAT
PSPC (MHR95).

For both A0620-00 and Cen~X-4, the X-ray data and the UV data were
obtained at different epochs. Thus, for Cen~X-4 we expect that source
variability introduces an uncertainty in this composite spectrum
(Fig.~4b) that is about a factor of 3 (see $\S3.2$), which is
comparable to the measurement uncertainties for the X-ray data (Asai
et al. 1998).  Source variability for A0620-00 appears to be less
($\S3.2$) and may affect the UV/X-ray spectrum by a factor
$\lesssim~2$, which is somewhat less than the uncertainty in the
measured X-ray flux (MHR95). Thus, the appearance of Figure~4ab is
unlikely to be affected significantly by source variability.

{\it A key result shown in Figure~4ab is the very different character of
the broadband spectra of the two sources in passing from the optical
to the X-ray: the flux ($\nu$F$_{\nu}$) of A0620-00 falls by a factor
of $\sim~70$, whereas, the flux of Cen~X-4 decreases by only a factor of
2-3.}

\section{Discussion}

A strong Mg~II $\lambda2800$ emission line is present in the spectra
of both A0620-00 and Cen~X-4 (Fig.~1; $\S3.1$).  Mg~II $\lambda2800$
is especially broad and intense in the spectrum of A0620-00, and it is
natural to compare it to H$_{\alpha}$ in emission, the line that
dominates the optical spectrum of A0620-00.  The average width of the
Mg~II line measured for the seven individual spectra is FWHM = 2500
$\pm$ 220 km s$^{-1}$, which is comparable to the width of
H$_{\alpha}$: FWHM $\approx 2000$ km s$^{-1}$ (Marsh et al. 1994;
Orosz et al.  1994).  However, the line profiles are very different.
The Mg~II line is plainly single peaked (Fig.~1a), given the adequate
spectral resolution of the STIS (3.2\AA\ or 340 km s$^{-1}$); on the
other hand, the H$_{\alpha}$ line is always double peaked.  The broad
and double-peaked H$_{\alpha}$ line is ubiquitous in the spectra of
quiescent X-ray novae and cataclymic variables, and is generally
attributed to emission from an accretion disk (e.g. Horne \& Marsh
1986).  We know of no ready explanation for the single-peaked profile
of the Mg~II line.

It is also interesting to consider for each system the luminosities of
the H$_{\alpha}$ and Mg~II lines relative to the X-ray luminosity.
For A0620-00 we find (in units of $10^{30}$ ergs~s$^{-1}$): L$_{\rm
H_{\alpha}} \approx 3.4$, L$_{\rm Mg~II} \approx 3.4$ and L$_{\rm x}
\approx 6$ (Oke 1977; Marsh et al. 1994; Orosz et al. 1994; MHR95).
Thus the line and X-ray luminosities are very comparable; plainly the X-ray
source is incapable of powering the lines.  For Cen X-4 we find:
L$_{\rm H_{\alpha}} \approx 1.2$, L$_{\rm Mg~II} \approx 2.8$ and
L$_{\rm x} \approx 250$ (Shahbaz et al. 1993; Shahbaz et al. 1996;
Asai et al. 1998).  The line luminosities for the two systems are
quite similar, although the X-ray luminosity of Cen X-4 is greater by
a factor of $\sim 40$.  For both systems, we conclude that both the
H$_{\alpha}$ and Mg~II lines are very probably produced in an outer
accretion disk or gas stream (Horne \& Marsh 1986) and are little
affected by the radiation from the central X-ray source.

As shown in Figure~4, the flux ($\nu$F$_{\nu}$) of A0620-00 falls by a
factor of $\approx 70$ in passing from the peak in the UV/optical band
to the X-ray band. Comparable data exist for only two other black-hole
X-ray novae, and in these cases the (dereddened) UV/optical flux is
also much greater than the X-ray flux: The flux of V404 Cyg decreases
by a factor of $\approx 20$ (Narayan et al. 1997a) and GRO J1655-40 by
a factor of $\approx 50$ (Hameury et al. 1997).

Among the black-hole X-ray novae, the break in the UV spectrum for
log($\nu$)~$\gtrsim~14.9$ is presently observable only for A0620-00
(Fig.~4a) because its IS extinction is moderate, the source is bright,
and the cool secondary does not contaminate the UV spectrum. FOS
observations of A0620-00 in HST Cycle 1 provided some preliminary
evidence for the existence of an optical/UV peak (MHR95; Narayan et
al. 1996); however, the results were inconclusive due to serious
difficulties in determining the background rate at short wavelengths
(MHR95).  Now, however, these earlier indications of a break in the
spectrum are amply confirmed by the STIS observations reported
here. Jointly the FOS and STIS observations establish a peak near
log($\nu) \sim 14.9$.  The flux is seen to fall abruptly by a factor
of $\approx 3$ in the UV band, and to be $\approx 70$ times fainter in
the X-ray band (Fig.~4a).

The striking UV/optical peak in the spectrum of A0620-00 (Fig.~4a) has
been ascribed in ADAF models to synchrotron emission (Narayan et
al. 1996; Narayan et al. 1997a).  Since synchrotron emission is a
strong function of electron temperature, the bulk of this emission
comes from within $\sim$~10 R$_{S}$ (R$_{S} = 2GM/c^{2}$) of the black
hole. ADAF models have yielded good fits to all of the available
broadband data (Narayan et al. 1997a; Hameury et al. 1997). That is,
the synchrotron peak in the models dominates over the X-ray flux by a
factor of $\sim 20-70$, as observed. Moreover, at higher energies the
observed photon index (0.7-8.5 keV) of V404 Cyg, $\Gamma =
2.1(+0.5,-0.3)$, is well matched by the Compton scattering and
bremsstrahlung components of the ADAF model spectrum (Narayan et
al. 1997a).  In these models, it was assumed that none of the
accreting mass was lost to a wind, and also that the fraction of the
turbulent energy that goes into heating the electrons is $\delta \sim
(m_{e}/m_{i}) \sim 10^{-3}$.

At the close of $\S1$, we listed several similarities between
A0620-00 and Cen~X-4.  Several dissimilarities have been pointed out
in the present work.  The broadband spectrum of Cen~X-4 (Fig.~4b) is
plainly different. First, compared to A0620-00 and other black-hole
X-ray novae, the decrease in the flux of Cen X-4 in going from the UV
band to the X-ray band is very small--a factor of $\sim 2$ (Fig.~4b).
Second, there is no peak in the {\it observed} UV/optical continuum;
the spectrum generally increases with increasing frequency.  An
observational challenge is to search for a break in the UV spectrum of
Cen X-4 to the shortest possible wavelengths.  This is important
since the peak of the thermal synchrotron emission in ADAFs depends on
the mass of the compact object:
$\nu_{s}~\sim~10^{15}$(M/M$_{\odot})^{-1/2}$ (Quataert \& Narayan
1999b).  Thus, in relation to the spectrum of A0620-00, the spectrum
of Cen X-4 may be expected to peak at wavelengths $\sim~1500$\AA\ (see
$\S1$). A corresponding theoretical challenge is to develop ADAF models for
quiescent accretion onto neutron stars (Yi et al. 1996).  This is a
difficult undertaking because a neutron star surface will re-radiate
the thermal energy it accretes via an ADAF, making it difficult to
observe the faint component of radiation due to the ADAF.  A black
hole, on the other hand, is much simpler because its event horizon
will hide the thermal energy (Narayan et al. 1997b).

From the observational point of view, compelling evidence now exists
for a large luminosity difference between black holes and neutron
stars at quiescent levels of accretion.  Chandra and XMM observations
that are now scheduled will soon provide conclusive results on this
luminosity divide.  In the context of the ADAF model, this luminosity
difference provides evidence that black holes possess an event horizon
(Narayan et al. 1997b; Garcia et al. 1998; Menou et al. 1999).  The
results presented here substantiate this earlier work and provide
additional information: A0620-00 and Cen X-4 are at very comparable
distances and have similar mass transfer rates ($\S1$).  Nevertheless,
A0620-00 is $\sim 40$ times less luminous than Cen X-4 in the X-ray
band ($\S1$), as expected if A0620-00 has an event horizon and Cen
X-4 does not.  At $\sim 1700$\AA\, however, A0620-00 is less
luminous by only a factor of $\sim 6$, and in the optical band, where
the observed luminosity is a maximum, the two systems are comparably
luminous (Fig.~4).

Recently, the ADAF model has been successfully applied to the spectra
of 10$^{8}-10^{10}M_{\odot}$ black holes in the nuclei of elliptical
galaxies (Di Matteo et al. 1999).  The observed spectra of most of
these supermassive black holes differ markedly from the spectra of
quiescent Galactic black holes in two respects.  First, the
synchrotron peak is dwarfed by the X-ray peak: For six galactic
nuclei, the peak synchrotron flux is typically only a few percent of
the peak X-ray flux (Di Matteo et al. 1999).  Second, their X-ray
spectra are significantly harder than the spectra of BHXN (photon
index $\Gamma = 0.6-1.5$; Allen, Di Matteo, \& Fabian 1999).
Consequently, the ADAF models used to fit the supermassive black holes
differ significantly from the models applied previously to the
Galactic black holes.  The fundamental difference in the new models is
the presence of a massive wind that can expel much or most of the
incoming accretion flow (Narayan \& Yi 1994, 1995; Blandford \&
Begelman 1999).  The newer models also consider much stronger heating
of the electrons ($\delta \sim 0.01-0.75)$ (Quataert \& Narayan 1999a,
1999b).

Because there is a degeneracy in the parameters, it has been possible
to construct new ADAF models for the X-ray nova V404 Cyg with strong
winds and large values of $\delta$ that are comparable in quality to
the earlier fits described above, which were achieved with no winds
and $\delta \sim 0.001$ (Quataert \& Narayan 1999b; Narayan et al. 1997a).
Observations with Chandra and XMM will soon tighten substantially the
constraints on the model fits.  Obviously, these observations will be
crucial to our explorations of both Galactic black holes and
supermassive black holes that are accreting via an ADAF. An important
goal is to obtain a unified ADAF model for low-luminosity black holes
with masses ranging from 10 M$_{\odot}$ to 10$^{10} M_{\odot}$.

\acknowledgements

We thank K. Menou, R. Narayan, E. Quataert, G. Sobczak and the
referee, K. Long, for helpful discussions and comments on the
manuscript; and L. Macri and E. Barton for making contemporaneous
V-band observations of our objects; and numerous people at STScI for
help with the STIS observations and data analysis, including S. Baum,
R. Bohlin, G. Kriss, H. Lanning, K. Peterson, K. Sahu and B. Simon.
Support for this work was provided by NASA through grant number
GO-07393.01-96A from the Space Telescope Science Institute, which is
operated by AURA, Inc., under NASA contract NAS5-26555.

\newpage
\begin{center}
\begin{tabular}{rccclcc}
\multicolumn{7}{c}{TABLE 1} \\
\multicolumn{7}{c}{JOURNAL OF OBSERVATIONS} \\
&&&&&& \\ \hline\hline
&\multicolumn{2}{c}{Start of
Exposure}&Object&Instrument&Wavelength&Obs.Time \\
 &\multicolumn{2}{c}{(1998 UT Date/Time)}&&&     Range (A)&     (s) \\
\hline

1 &  Mar 3 & 04:17:00&    A0620-00   &FLWO 1.2m&       V        &  900
\\

2\rlap{$^{\ast}$}  & Mar 4 & 05:35:03 &   A0620-00   &HST/STIS  &  1700-3200
&  2150
\\

3  & Mar 4 & 06:67:55  &  A0620-00   &HST/STIS   & 1700-3200  &   2550
\\

4  & Mar 4 & 08:35:06   & A0620-00   &HST/STIS    &1700-3200 &    2550
\\

5 &  Mar 4 & 10:11:55    &A0620-00   &HST/STIS&    1700-3200&     2550
\\

6  & Mar 4 & 11:48:43&    A0620-00   &HST/STIS &   1700-3200     &2550
\\

7  & Apr 1 & 09:43:15 &   Cen X-4    &FLWO 1.2m &      V        &  900
\\

8  & Apr 1 & 21:08:13  &  Cen X-4    &HST/STIS   & 1700-3200   &  2150
\\

9  & May 5 & 03:09:01   & A0620-00  & HST/STIS    &1700-3200  &   2150
\\

10 & May 5 & 04:40:25    &A0620-00 &  HST/STIS  &  1700-3200 &    2550
\\

11 & May 5&  06:17:35    &A0620-00&   HST/STIS   & 1700-3200&     2550
\\ \hline
\multicolumn{7}{p{6in}}{$^{\ast}$Spectrum not used in the present work
because its ``pipeline'' background is a factor of $\approx 3$ higher than
for all the other spectra.}
\end{tabular}
\end{center}

\newpage \figcaption[fig1.eps]{The summed spectrum of (a) A0620-00
(T$_{exp}=4.8$ hrs) and the spectrum of (b) Cen~X-4 (T$_{exp} = 0.6$
hrs). The spectra displayed here are binned at 2.0\AA\ and are not
corrected for the effects of IS reddening. The strong line in both
spectra is due to Mg~II.  Note that for A0620-00, the line is both
very broad (2500 km s$^{-1}$) and single peaked.}

\figcaption[fig2.eps]{ A comparison of summed STIS spectra obtained
two months apart.  The change in intensity is modest.  However, the
errors have been reliably and conservatively determined, and we
conclude that the continuum flux did decrease by a factor of $\approx
1.2-1.3$ during this two month period.  The spectra have not been
corrected for reddening.}

\figcaption[fig3.eps]{The UV/optical continuum spectra of A0620-00 and
Cen~X-4.  (a) Spectrum of A0620-00: The STIS data are plotted here as
filled circles; each error bar is computed from the standard deviation
of seven flux measurements.  The data have been corrected for a
reddening of E$_{B-V} = 0.35$. The STIS data are binned in 100\AA\
intervals with two exceptions: The highest frequency point is a
2$\sigma$ upper limit averaged over a 300\AA\ band, and the adjacent
data point is averaged over a 200\AA\ band.  The lower/upper
histograms represent the spectrum if one adopts a reddening of
E$_{B-V} = 0.30/0.40$, rather than the nominal value of 0.35.  The
1992 FOS data (open squares) are taken from Table~1 in NMY96. An
alternative optical spectrum of the non-stellar light, which was
derived by Marsh et al. (1994), is indicated by the dashed line (see
text). (b) Spectrum of Cen~X-4: The STIS UV data are plotted as before
for A0620-00.  Here, however, the errors are those due to counting
statistics (arbitrarily) multiplied by 3.0.  The spectrum has been
corrected for a nominal reddening of E$_{B-V} = 0.10$.  The
lower/upper histograms show the spectrum if one adopts a reddening of
0.05/0.15.}  

\figcaption[fig4.eps]{Broadband spectra of A0620-00 and Cen~X-4.  The
UV/optical data for both sources is the same as that shown in Figure~3.
The EUV and X-ray data are discussed in the text.
The ASCA X-ray upper limit in Figure~4a is at the $3\sigma$ level of
confidence.}
 
\newpage
\begin{figure}
\figurenum{1}
\plotone{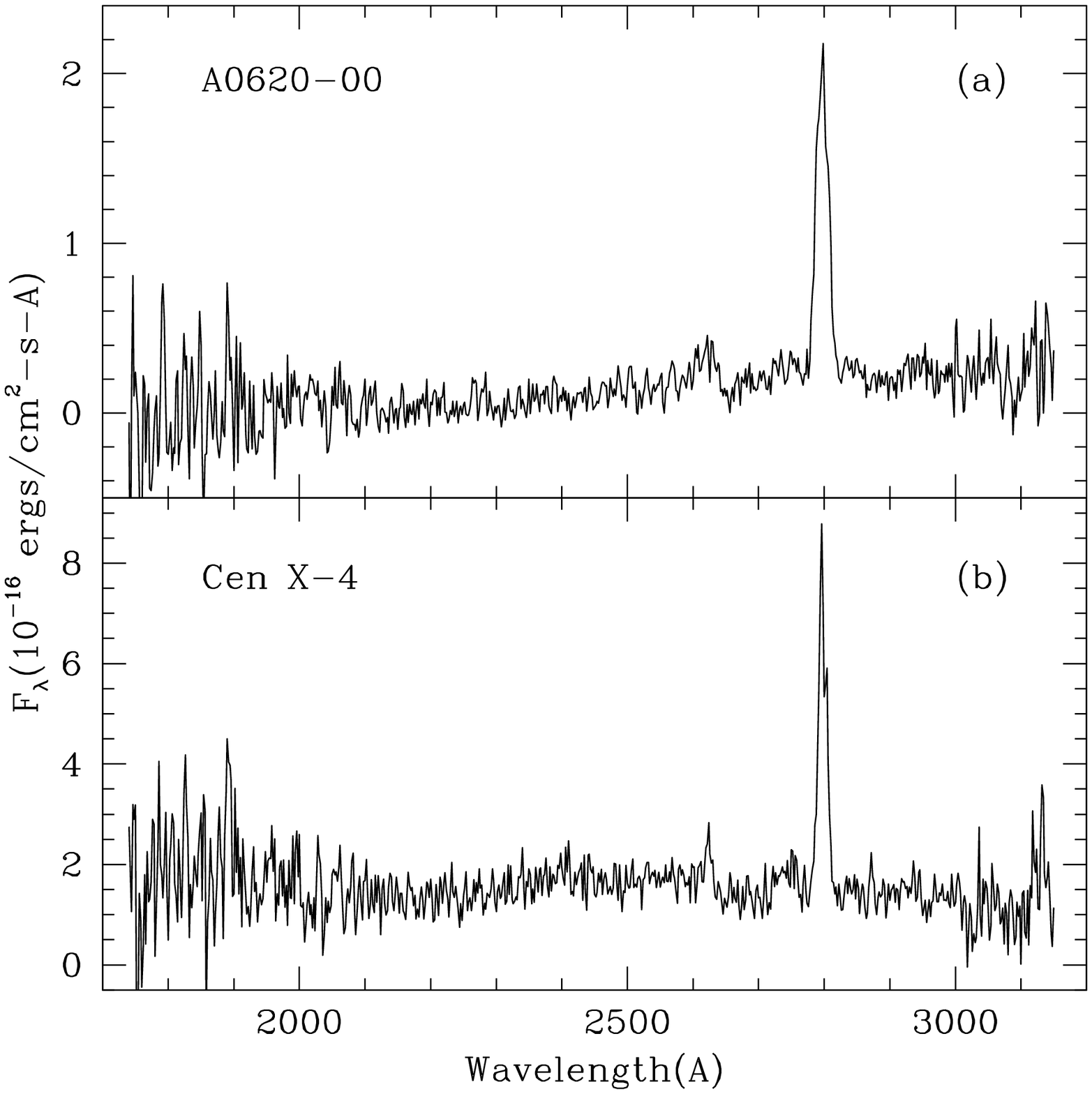}
\caption{ }
\end{figure}

\newpage
\begin{figure}
\figurenum{2}
\plotone{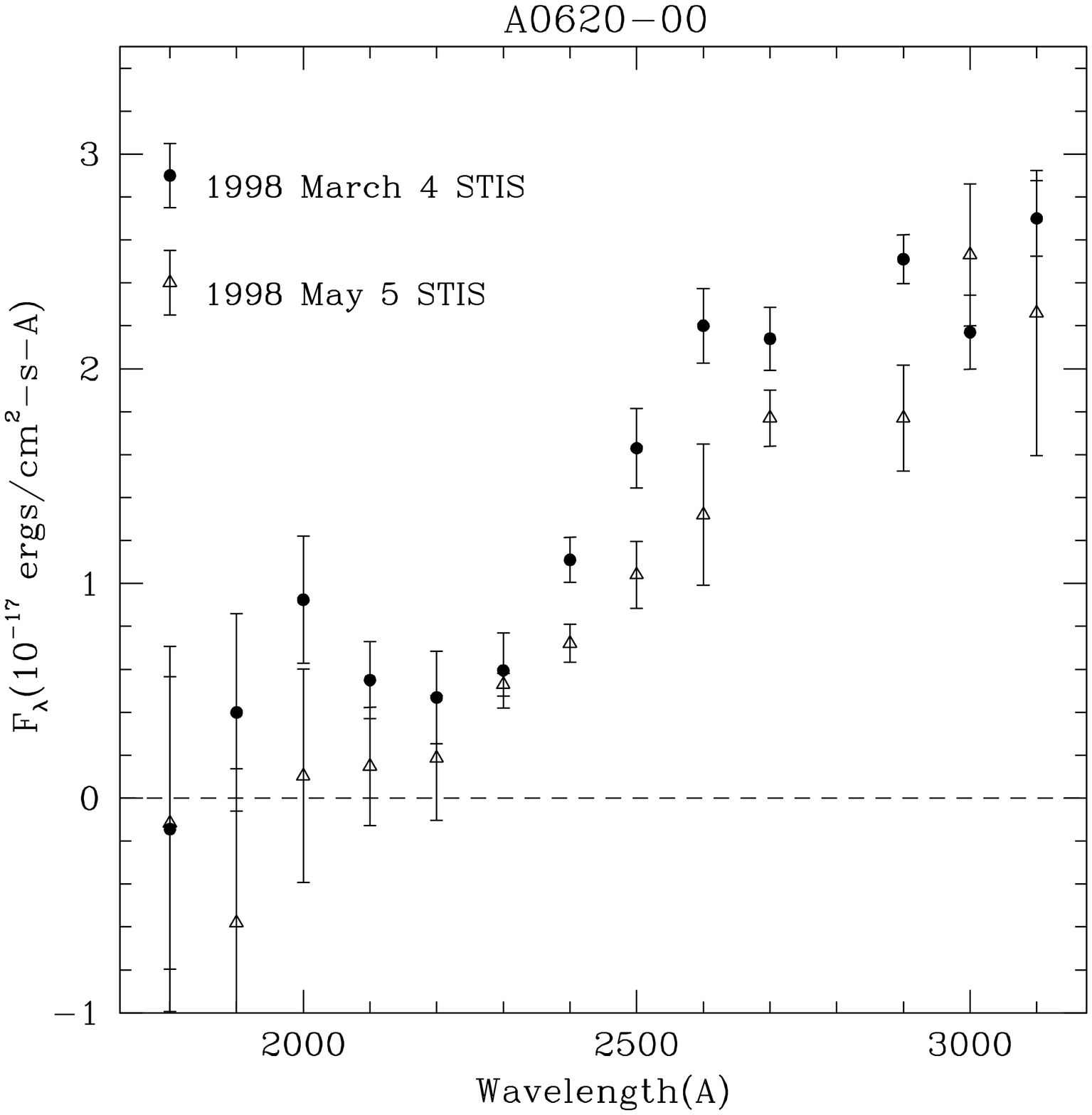}
\caption{ }
\end{figure}

\newpage
\begin{figure}
\figurenum{3}
\plotone{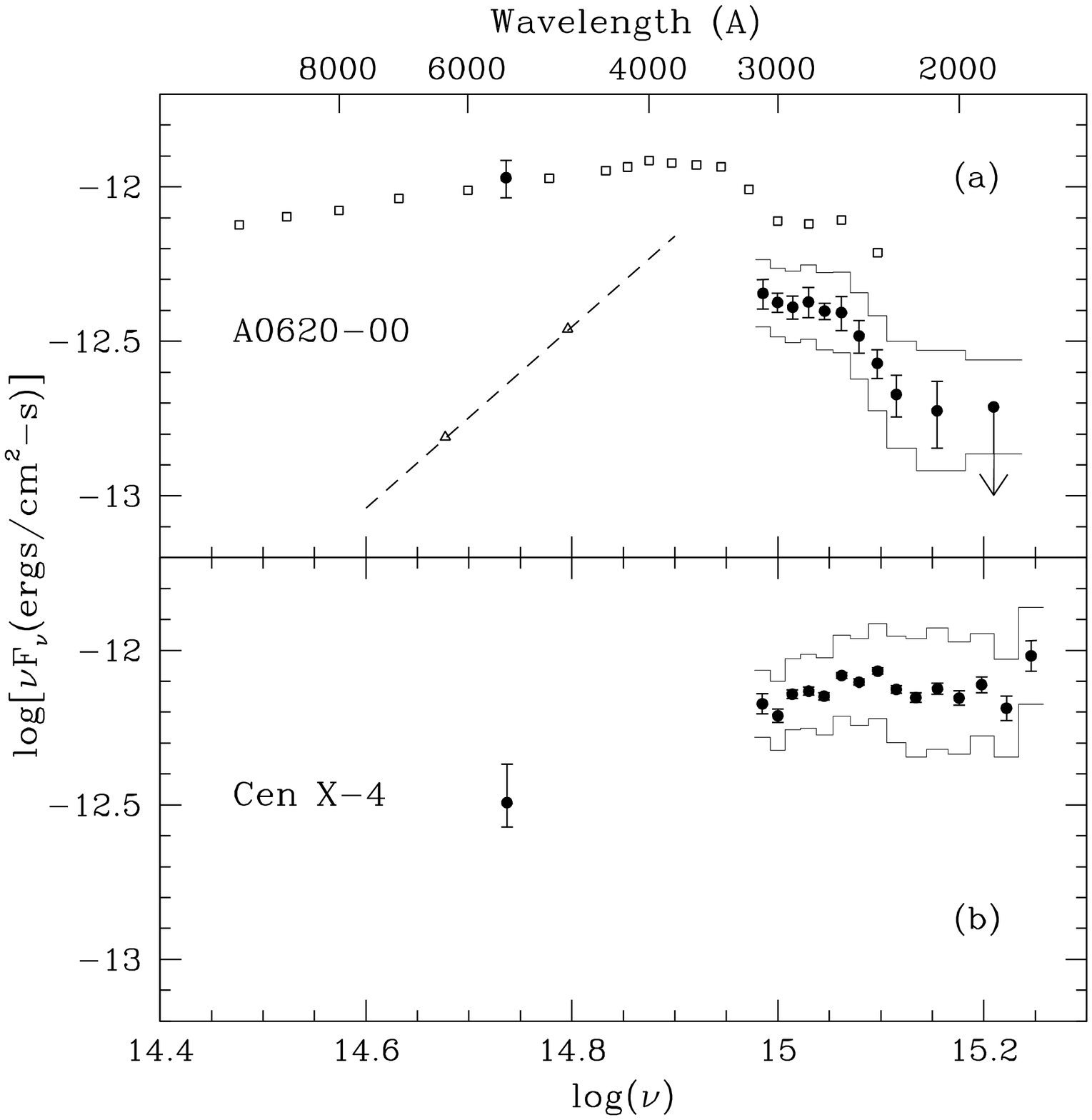}
\caption{ }
\end{figure}

\newpage
\begin{figure}
\figurenum{4}
\plotone{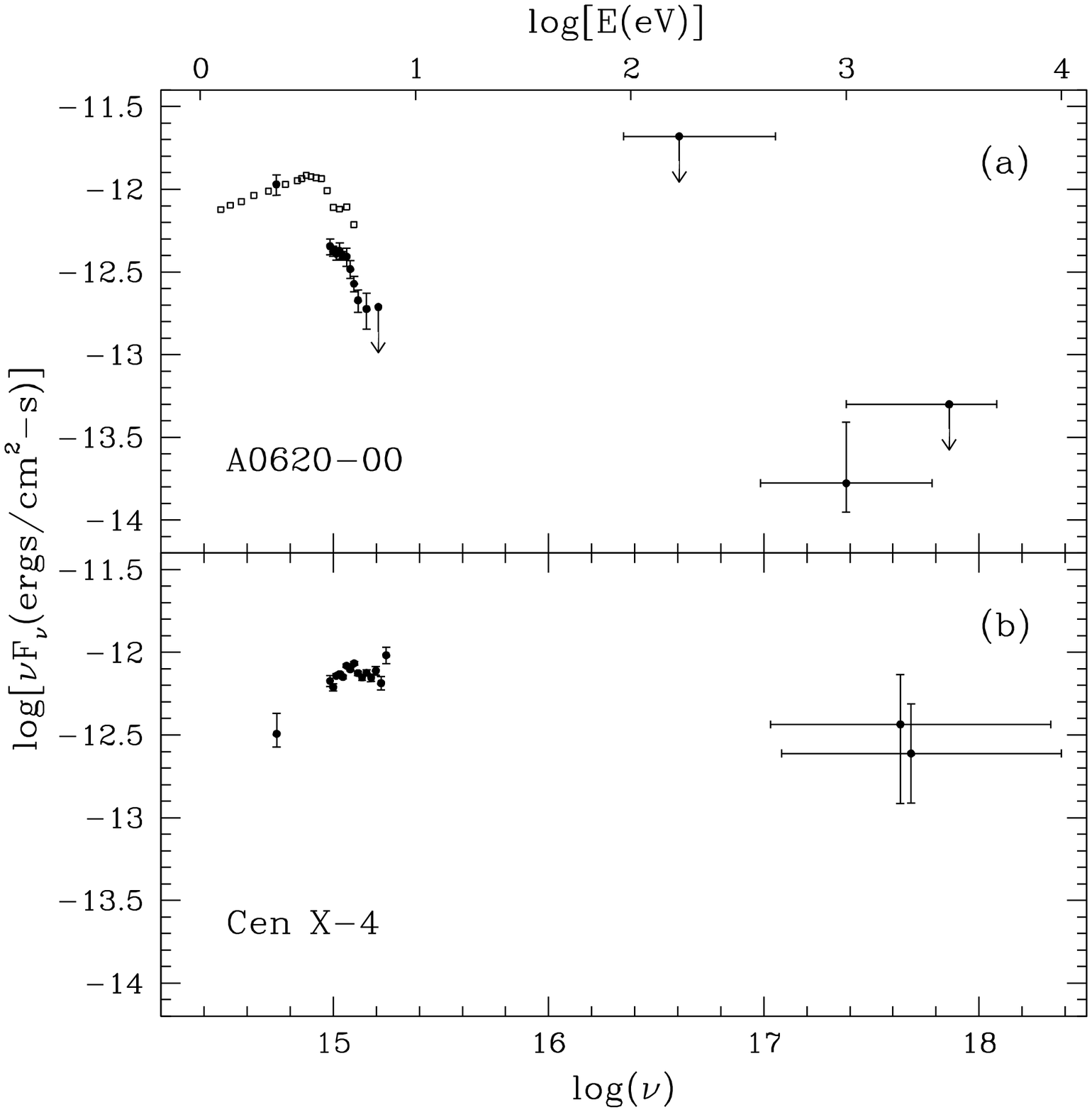}
\caption{ }
\end{figure}


\begin{references}

\reference{}Allen, S. W., Di Matteo, T., \& Fabian, A. C., 1999, \mnras,
submitted (astro-ph/9905052)

\reference{}Asai, K., Dotani, T., Hoshi, R., Tanaka, Y., Robinson,
C. R., \& Terada, K. 1998, \pasj, 50, 611

\reference{}Barret, D., McClintock, J. E., \& Grindlay, J. E. 1996,
\apj, 473, 963

\reference{}Blair, W. P., Raymond, J. C., Dupree, A. K., Wu, C.-C.,
Holm, A. V., \& Swank, J. H. 1984,  \apj,  278,  270

\reference{}Blandford, R. D., \& Begelman, M. C. 1999, \mnras, 303, L1

%\reference{}Campana, S., Colpi, M., Mereghetti, S., Stella, L.,
%Tavani, M. 1998, A\&A Rev., 8, 279

\reference{}Campana, S., Mereghetti, S., Stella, L., \& Colpi,
M. 1997, A\&A, 324, 941

\reference{}Cardelli, J. A., Clayton, G. C., \& Mathis, J. S. 1989,
\apj, 345, 245

\reference{}Chevalier, C., Ilovaisky, S. A., van Paradijs, J.,
Pedersen, H., \& van der Klis, M. 1989, A\&A, 210, 114

\reference{}Cowley, A. P., Hutchings, J. B., Schmidtke, P. C., 
Hartwick, F. D. A., Crampton, D., \& Thompson, I. B. 1988, \aj, 95, 1231

\reference{}de Kool, M. 1988, \apj, 334, 336

\reference{}Di Matteo, T., Quataert, E., Allen, S. W., Narayan, R., \&
Fabian, A. C. 1999, \mnras, submitted (astro-ph/9905053)

\reference{}Garcia, M. R., McClintock, J. E., Narayan, R., \&
Callanan, P. 1998, in the Proceedings of the 13th North American
Workshop on CVs, eds. S. Howell, E. Kuulkers, \& C. Woodward, p. 506
(astro-ph/9708149)

\reference{}Horne, K., \& Marsh, T. R. 1986, \mnras, 218, 761

\reference{}Huang, M., \& Wheeler, J. C. 1989, \apj, 343, 229

\reference{}Landolt, A. U. 1992, \aj, 104, 340 

\reference{}Long, K. S., Helfand, D. J., \& Grabelsky, D. A. 1981,
\apj, 248, 925

\reference{}Hameury, J.-M., Lasota, J.-P., McClintock, J.E., \&
Narayan, R. 1997, \apj, 489, 234

\reference{}Marsh, T. R., Robinson, E. L., \& Wood, J. H. 1994,
\mnras, 266, 137

\reference{}Menou, K., Esin, A. A., Narayan, R., Garcia, M. R., Lasota, 
J.-P., \& McClintock, J. E. 1999, \apj, 520, 276

\reference{}McClintock, J. E. 1986, in The Physics of Accretion onto
Compact Objects, ed.\ K. O. Mason, M. G. Watson, \& N. E. White
(Berlin: Springer), 211

\reference{}McClintock, J. E., Horne, K., \& Remillard, R. A. 1995,
\apj, 442, 358 (MHR95)

\reference{}McClintock, J. E., Petro, L. D., Remillard, R. A., \&
Ricker, G. R.  1983, \apjl, 266, L27

\reference{}McClintock, J. E., \& Remillard, R. A. 1986, \apj, 308,
110

\reference{}McClintock, J. E., \& Remillard, R. A. 1990, \apj, 350, 
386

\reference{}Narayan, R., \& Yi, I. 1994, \apjl, 428, L13

\reference{}Narayan, R., \& Yi, I. 1995, \apj, 452, 710

\reference{}Narayan, R., Barret, D., \& McClintock, J. E. 1997a,
\apj, 482, 448

\reference{}Narayan, R., Garcia, M. R., \& McClintock, J. E. 1997b,
\apj, 478, L79

\reference{}Narayan, R., McClintock, J. E., \& Yi, I. 1996, \apj,
457, 821

\reference{}Oke, J. B. 1977, ApJ, 217, 181

\reference{}Oke, J. B., \& Greenstein, J. L. 1977, \apj, 211, 872

\reference{}Orosz, J. A., Bailyn, C. D., Remillard, R. A., 
McClintock, J. E., \& Foltz, C. B. 1994, \apj, 436, 848

\reference{}Quataert, E. \& Narayan, R. 1999a, \apj, 516, 399

\reference{}Quataert, E. \& Narayan, R. 1999b, \apj, 520, 298

\reference{}Shahbaz, T., Naylor, T., \& Charles, P. A. 1993, \mnras,
265, 655

\reference{}Shahbaz, T., Naylor, T., \& Charles, P. A. 1994, \mnras,
268, 756

\reference{}Shahbaz, T., Smale, A. P., Naylor, T., Charles, P. A., van
Paradijs, J., Hassall, B. J. M., \& Callanan, P. 1996, \mnras, 282,
1437

\reference{}Wu, C.-C., Aalders, J. W. G., van Duinen, R. J., Kester, D., 
\& Wesselius, P. R. 1976, A\&A, 50, 445

\reference{}Wu, C.-C., Panek, R. J., Holm, A. V., Schmitz, M., \&
Swank, J. H. 1983, \pasp, 95, 391

\reference{}Yi, I., Narayan, R., Barret, D., \& McClintock, J. E. 
1996, A\&A Supp., 120, 187

\end{references}
\end{document}